\documentclass[prd,%
preprint,%
tightenlines,showpacs,%
nofootinbib,%
amsfonts,amsmath]{revtex4}

\begin{document}

\title{Covariant Galileon}

\author{C.~Deffayet} \email{deffayet@iap.fr}
\affiliation{AstroParticule \& Cosmologie,
UMR 7164-CNRS, Universit\'e Denis Diderot-Paris 7,
CEA, Observatoire de Paris,
10 rue Alice Domon et L\'eonie
Duquet, F-75205 Paris Cedex 13, France}

\author{G.~\surname{Esposito-Far\`ese}} \email{gef@iap.fr}
\affiliation{${\mathcal{G}}{\mathbb{R}}
\varepsilon{\mathbb{C}}{\mathcal{O}}$, Institut d'Astrophysique
de Paris, UMR 7095-CNRS, Universit\'e Pierre et Marie
Curie-Paris 6, 98bis boulevard Arago, F-75014 Paris, France}

\author{A.~Vikman} \email{alexander.vikman@nyu.edu}
\affiliation{{}CCPP, New York University,
Meyer Hall of Physics, 4 Washington Place,
New York, NY 10003,USA}

\begin{abstract}
We consider the recently introduced ``galileon'' field in a
dynamical spacetime. When the galileon is assumed to be minimally
coupled to the metric, we underline that both field equations of
the galileon and the metric involve up to third-order
derivatives. We show that a unique nonminimal coupling of the
galileon to curvature eliminates all higher derivatives in all
field equations, hence yielding second-order equations, without
any extra propagating degree of freedom. The resulting theory
breaks the generalized ``Galilean'' invariance of the original
model.
\end{abstract}

\date{January 9, 2009}


\maketitle

\section{Introduction}
An interesting scalar field theory, named {\it galileon}, was
recently introduced in Ref.~\cite{Nicolis:2008in} inspired by the
Dvali-Gabadadze-Porrati (DGP) model \cite{Dvali:2000hr} and its
ability to produce an accelerated expansion of the Universe
without introducing any dark energy nor cosmological constant
\cite{Deffayet:2000uy,Deffayet:2001pu}. More specifically, taking
the so-called decoupling limit of the DGP model, one can extract
an effective theory for a scalar field $\pi$ argued to describe
the scalar sector of the original model
\cite{Luty:2003vm,Nicolis:2004qq}. In particular, this scalar
field theory still exhibits interesting properties of the
original model, such as the existence of the self-accelerating
branch of DGP cosmology. Similarly to the decoupling limit of
massive gravity \cite{ArkaniHamed:2002sp}, the action of this
effective theory contains second-order derivatives acting on the
scalar field. But in contrast to massive gravity, in which the
equations of motion contain fourth-order derivatives, signaling
the presence of ghost-like modes or at least extra degrees of
freedom \cite{Deffayet:2005ys,Creminelli:2005qk}, the field
equation for $\pi$ is in fact of {\it second} order. Indeed, it
is well known that, according to Ostrogradski's theorem
\cite{Ostrogradski}, higher-derivative theories contain extra
degrees of freedom, and are usually plagued by negative energies
and related instabilities (see e.g. \cite{Woodard:2006nt}). In
the decoupling limit of the DGP model\footnote{Note that this
decoupling limit not only exists in massive gravity and in the
DGP model, but also in the generalized framework of
Ref.~\cite{DEGRAV}.}, it is thus striking that such additional
degrees of freedom do not appear. Moreover, in addition to the
usual constant-shift symmety $\pi \rightarrow \pi+c$ in field
space, due to the absence of undifferentiated $\pi$'s in the
action, this theory also possesses a symmetry under constant
shifts of the gradient $\partial_{\mu} \pi \rightarrow
\partial_{\mu} \pi + b_{\mu}$. This symmetry implies that the
$\pi$ field equation contains \textit{only} second derivatives
(but no first derivative nor any undifferentiated field).

The recent reference \cite{Nicolis:2008in} classified all
possible 4-dimensional actions for a scalar field, $\pi$, which
have the same properties as the DGP effective theory discussed
above: having Lorentz invariant equations of motion which contain
only second derivatives of $\pi$ on a flat (Minkowski)
background. Besides known Lagrangians which are (i)~linear in
$\pi$, ${\cal L}_1=\pi$, (ii)~the standard quadratic kinetic
Lagrangian for $\pi$ in the form ${\cal L}_2=\partial_\mu \pi
\partial^\mu \pi$ and (iii)~a cubic Lagrangian ${\cal L}_3=\Box
\pi \partial_\mu \pi \partial^\mu \pi$ which is the one obtained
in the decoupling limit of DGP model, Ref.~\cite{Nicolis:2008in}
argues there only exist two possible other Lagrangians which
share the same property in 4 spacetime dimensions. The first one,
named ${\cal L}_4$ is made of a linear combination of four terms
each made of a product of four $\pi$ and a total of six
derivatives acting on $\pi$. The second one, named ${\cal L}_5$
is made of a linear combination of seven terms each made of a
product of five $\pi$ and a total of eight derivatives acting on
$\pi$. The linear combinations in ${\cal L}_4$ and ${\cal L}_5$
are uniquely chosen (up to total derivatives) so that the
equations of motions derived from the corresponding action only
contain second-order derivatives acting on $\pi$. The analysis
carried in \cite{Nicolis:2008in} was made in flat spacetime and
one can legitimately expect things to change radically when one
considers the same theory on a curved dynamical spacetime.

Indeed, first, when varied with respect to $\pi$, the Lagrangians
${\cal L}_4$ and ${\cal L}_5$ are expected to generate
third-order derivatives acting on the metric. These derivatives,
in the form of gradients of the Riemann tensor, appear via
commutations of fourth-order covariant derivatives acting on
$\pi$, which eventually disappear (together with third-order
derivatives). Second, when varied with respect to the metric, the
same Lagrangians are expected to generate third-order derivatives
acting on $\pi$, from the covariant derivatives present in the
$\pi$ action. Should such higher derivatives be generated,
instabilities are expected to be present, or at least extra
degrees of freedom to propagate, as can simply be understood from
the counting of initial conditions necessary to have a well posed
Cauchy problem. In fact, it is argued in \cite{Nicolis:2008in}
that the galileon model cannot be covariantized while keeping all
its desired properties, in particular its symmetries and the
absence of ghosts.

It is our purpose here to examine this question in some detail.
We will first show that the naive expectation summarized above
turns out to be true: namely a naive covariantization of the
$\pi$ action leads to third-order derivatives both in the $\pi$
and the metric field equations. However, we will show that there
is a possible (and unique\footnote{This nonminimal coupling is
unique up to total derivatives, and provided all $\pi$'s are
differentiated.}) nonminimal coupling to curvature that has the
property of removing at the same time higher derivatives in both
equations of motions (those of $\pi$ and in the $\pi$
energy-momentum tensor). The full set of equations concerning the
simpler Lagrangian ${\cal L}_4$ will be discussed in the bulk of
the paper, while the heaviest ones corresponding to the more
complex ${\cal L}_5$ will be given in an appendix.

\section{Minimally vs nonminimally-coupled galileon}
As shown in \cite{Nicolis:2008in}, any linear combination of
${\cal L}_1$, ${\cal L}_2$, ${\cal L}_3$, defined in the
introduction, ${\cal L}_4$ and ${\cal L}_5$, defined as
follows\footnote{We use the sign conventions of Ref.~\cite{MTW},
notably the mostly-plus signature.}
\begin{eqnarray}
{\cal L}_4 &=& \left(\Box \pi\right)^2 \left(\pi_{;\mu}\,\pi^{;\mu}\right)
- 2 \left(\Box \pi\right)\left(\pi_{;\mu}\,\pi^{;\mu\nu}\,\pi_{;\nu}\right)
\nonumber \\
&& - \left(\pi_{;\mu\nu}\,\pi^{;\mu\nu}\right) \left(\pi_{;\rho}\,\pi^{;\rho}\right)
+2 \left(\pi_{;\mu}\pi^{;\mu\nu}\,\pi_{;\nu\rho}\,\pi^{;\rho}\right), \label{L4}\\
{\cal L}_5 &=& \left(\Box \pi\right)^3 \left(\pi_{;\mu}\,\pi^{;\mu}\right)
- 3 \left(\Box \pi\right)^2\left(\pi_{;\mu}\,\pi^{;\mu\nu}\,\pi_{;\nu}\right)
- 3 \left(\Box \pi\right) \left(\pi_{;\mu\nu}\,\pi^{;\mu\nu}\right) \left(\pi_{;\rho}\,\pi^{;\rho}\right)
\nonumber \\
&& +6 \left(\Box \pi\right)\left(\pi_{;\mu}\pi^{;\mu\nu}\,\pi_{;\nu\rho}\,\pi^{;\rho}\right)
+2 \left(\pi_{;\mu}^{\hphantom{;\mu}\nu}\,\pi_{;\nu}^{\hphantom{;\nu}\rho}\,\pi_{;\rho}^{\hphantom{;\rho}\mu}\right) \left(\pi_{;\lambda}\,\pi^{;\lambda}\right) \nonumber \\
&& +3 \left(\pi_{;\mu\nu}\,\pi^{;\mu\nu}\right)
\left(\pi_{;\rho}\,\pi^{;\rho\lambda}\,\pi_{;\lambda}\right)
-6 \left(\pi_{;\mu}\,\pi^{;\mu\nu}\,\pi_{;\nu\rho}\,\pi^{;\rho\lambda}\,\pi_{;\lambda}\right), \label{L5}
\end{eqnarray}
(where a semicolon denotes the covariant derivative $\nabla_\mu$
associated with the metric $g_{\mu \nu}$ and $\pi$ is a scalar
field, the galileon) has the property that, considered on flat
spacetime where $\nabla_\mu = \partial_\mu$, the derived
equations of motion for $\pi$ only contain second-order
derivatives of $\pi$. Actually, the various terms written in
Eqs.~(\ref{L4}) and (\ref{L5}) are not independent in flat
spacetime. For instance, the combination
$\left(\Box \pi\right)^2 \left(\pi_{;\mu}\,\pi^{;\mu}\right)
+ 2 \left(\Box \pi\right)\left(\pi_{;\mu}\,\pi^{;\mu\nu}\,\pi_{;\nu}\right)
- \left(\pi_{;\mu\nu}\,\pi^{;\mu\nu}\right) \left(\pi_{;\rho}\,\pi^{;\rho}\right)
-2 \left(\pi_{;\mu}\pi^{;\mu\nu}\,\pi_{;\nu\rho}\,\pi^{;\rho}\right)$
is a total derivative in Minkowski spacetime, so that ${\cal
L}_4$ may be rewritten more simply as
${\cal L}_4 = 2 \left(\pi_{;\rho}\,\pi^{;\rho}\right)
\left[\left(\Box \pi\right)^2 - \left(\pi_{;\mu\nu}\,\pi^{;\mu\nu}\right)\right]
+ \text{tot. div.}$
However, this expression does differ from Eq.~(\ref{L4}) in
curved background, and we will come back to this below. Similar
rewritings also exist for Eq.~(\ref{L5}) in Minkowski spacetime,
while giving different expressions in curved background.

Following Ref. \cite{Nicolis:2008in} a minimal coupling to the
metric and matter of the galileon could be defined by the action
\begin{eqnarray} \label{Sm}
S=\int d^4 x \sqrt{-g} \left(R + {\cal L}_\pi
+ {\cal L}_\text{matter}\right),
\end{eqnarray}
where ${\cal L}_\pi$ is a linear combination (with arbitrary
constant coefficients $c_i$) of ${\cal L}_1$, ${\cal L}_2$,
${\cal L}_3$, ${\cal L}_4$ and ${\cal L}_5$
\begin{eqnarray}
{\cal L}_\pi = \sum_{i=1}^{i=5} c_i\, {\cal L}_i,
\end{eqnarray}
$R$ is the Ricci scalar for the metric $g_{\mu\nu}$,
and ${\cal L}_\text{matter}$ is the Lagrangian of matter fields
minimally coupled to a metric $\tilde{g}_{\mu \nu}$ made of the
Einstein frame metric $g_{\mu \nu}$ and of the galileon $\pi$,
e.g. in a conformal way like in $\tilde{g}_{\mu \nu} = e^{2\pi}
g_{\mu \nu}$. It is clear that neither ${\cal L}_1$ nor
${\cal L}_2$ are able to generate equations of motion containing
derivatives of order higher than two when varied with respect to
$\pi$ or $g_{\mu \nu}$. The DGP-motivated Lagrangian ${\cal L}_3$
is known to generate a second-order field equation for $\pi$, and
it cannot yield either higher derivatives when varied with
respect to the metric. Indeed, it contains at most first
derivatives of $g_{\mu \nu}$, and those are multiplied by first
derivatives of $\pi$. As we will see now, the situation is
strikingly different for ${\cal L}_4$ and ${\cal L}_5$.

Let us first discuss the case of ${\cal L}_4$. When one varies
${\cal L}_4$ with respect to $\pi$, we obtain the equation of
motion ${\cal E}_4=0$, where\footnote{Equation~(\ref{E4a})
may also be written in a slightly different form by using the
identity
$\pi^{;\mu}\,\pi^{;\nu} \left(\pi_{;\mu\rho\nu}^{\hphantom{;\mu\rho\nu}\rho}
-\pi_{;\mu\rho\hphantom{\rho}\nu}^{\hphantom{;\mu\rho}\rho}\right)
+ \pi_{;\mu}\,\pi^{;\mu\nu}\left(\pi_{;\rho\hphantom{\rho}\nu}^{\hphantom{;\rho}\rho}- \pi_{;\nu\rho}^{\hphantom{;\nu\rho}\rho}\right)
- \pi^{;\mu}\, \pi^{;\nu\rho} \left( \pi_{;\nu\rho\mu} - \pi_{;\mu\nu\rho} \right)=0$.}
\begin{eqnarray}
{\cal E}_4&\equiv& 2\left(\pi_{;\mu}\,\pi^{;\mu}\right)\left(\pi_{;\nu\hphantom{\nu}\rho}^{\hphantom{;\nu}\nu\hphantom{\rho}\rho}
- \pi_{;\nu\rho}^{\hphantom{;\nu\rho}\nu\rho} \right)
+ 2\, \pi^{;\mu}\,\pi^{;\nu} \left(2\, \pi_{;\mu\rho\nu}^{\hphantom{;\mu\rho\nu}\rho} - \pi_{;\mu\nu\rho}^{\hphantom{;\mu\nu\rho}\rho} - \pi_{;\rho\hphantom{\rho}\mu\nu}^{\hphantom{;\rho}\rho}\right) \nonumber \\
&&+10 \left(\Box \pi \right) \pi^{;\mu}\left(\pi_{;\mu\nu}^{\hphantom{;\mu\nu}\nu}
- \pi_{;\nu\hphantom{\nu}\mu}^{\hphantom{;\nu}\nu}\right)
+ 12\, \pi_{;\mu}\,\pi^{;\mu\nu}\left(\pi_{;\rho\hphantom{\rho}\nu}^{\hphantom{;\rho}\rho}- \pi_{;\nu\rho}^{\hphantom{;\nu\rho}\rho}\right)
\nonumber \\
&&+ 8\, \pi^{;\mu}\, \pi^{;\nu\rho} \left( \pi_{;\nu\rho\mu} - \pi_{;\mu\nu\rho} \right)\nonumber \\
&&- 4 \left(\Box \pi \right)^3
-8 \left(\pi_{;\mu}^{\hphantom{;\mu}\nu}\,\pi_{;\nu}^{\hphantom{;\nu}\rho}\,\pi_{;\rho}^{\hphantom{;\rho}\mu}\right)
+12 \left(\Box \pi\right) \left(\pi_{;\mu\nu}\, \pi^{;\mu\nu} \right),
\label{E4a}
\end{eqnarray}
where the first two terms contain fourth-order derivatives, the
following three terms contain third-order derivatives and the
last three terms contain second-order derivatives. One notices in
fact that the fourth and third-order derivatives disappear on a
flat spacetime (as they should according to Ref.
\cite{Nicolis:2008in}). Indeed, commuting the derivatives, we
find that one can rewrite ${\cal E}_4$ as
\begin{eqnarray}
{\cal E}_4&=& - 4 \left(\Box \pi \right)^3
-8 \left(\pi_{;\mu}^{\hphantom{;\mu}\nu}\,\pi_{;\nu}^{\hphantom{;\nu}\rho}\,\pi_{;\rho}^{\hphantom{;\rho}\mu}\right)
+12 \left(\Box \pi\right) \left(\pi_{;\mu\nu}\pi^{;\mu\nu} \right)
- \left(\pi_{;\mu}\,\pi^{;\mu} \right)\left(\pi_{;\nu} \,R^{;\nu}\right) \nonumber \\
&&+2 \left(\pi_{;\mu}\,\pi_{;\nu}\,\pi_{;\rho} R^{\mu \nu;\rho}\right)
+10 \left(\Box \pi\right)\left(\pi_{;\mu}\,R^{\mu \nu}\,\pi_{;\nu}\right)
- 8 \left(\pi_{;\mu}\,\pi^{;\mu\nu}\,R_{\nu\rho}\,\pi^{;\rho}\right) \nonumber \\&&
- 2 \left(\pi_{;\mu}\,\pi^{;\mu}\right) \left(\pi_{;\nu\rho}\,R^{\nu \rho} \right)
- 8 \left(\pi_{;\mu}\,\pi_{;\nu}\,\pi_{;\rho\sigma}\,R^{\mu \rho \nu \sigma}\right).
\end{eqnarray}
We are left over with derivatives of the Ricci tensor and scalar
and hence with third-order derivatives of the metric. One can
think of a nonminimal coupling to the metric which would get rid
of those terms, in a form of a linear combination of the two
terms ${\cal L}_{4,1}$ and ${\cal L}_{4,2}$ defined
as\footnote{These terms are the only possible nonvanishing
ones made of contractions of four gradients of $\pi$ and one
curvature tensor, thereby involving a total of 6 derivatives.}
\begin{subequations}
\begin{eqnarray} \label{R1}
{\cal L}_{4,1}&=& \left(\pi_{;\mu}\,\pi^{;\mu}\right) \left(\pi_{;\nu}\,\pi^{;\nu}\right) R,\\
\label{R2}
{\cal L}_{4,2}&=&\left(\pi_{;\lambda}\,\pi^{;\lambda}\right) \left(\pi_{;\mu} \,R^{\mu \nu}\,\pi_{;\nu}\right).
\end{eqnarray}
\end{subequations}
In fact there is a unique combination of those two terms, namely
${\cal L}_{4,2}-\frac{1}{2}{\cal L}_{4,1}$ which added to ${\cal
L}_4$ eliminates all the third derivatives in the $\pi$ equations
of motion. Specifically, if we add to action
\begin{eqnarray} \label{S4}
S_4=\int d^4 x \sqrt{-g}\, {\cal L}_4
\end{eqnarray}
the action
\begin{eqnarray} \label{Snm4}
S^\text{nonmin}_4&\equiv& \int d^4 x \sqrt{-g} \left(\pi_{;\lambda}\,\pi^{;\lambda}\right)
\pi_{;\mu}\Bigl[R^{\mu \nu} - \frac{1}{2} g^{\mu \nu} R\Bigr] \pi_{;\nu}
\nonumber\\
&=& \int d^4 x \sqrt{-g} \left(\pi_{;\lambda}\,\pi^{;\lambda}\right)
\left(\pi_{;\mu}\,G^{\mu \nu}\,\pi_{;\nu}\right),
\end{eqnarray}
$G^{\mu \nu}$ denoting the Einstein tensor,
we obtain the equations of motion for $\pi$ in the form
${\cal E}'_4 = 0$, where ${\cal E}'_4$ is given by
\begin{eqnarray}
{\cal E}'_4 &=& - 4 \left(\Box \pi \right)^3
-8 \left(\pi_{;\mu}^{\hphantom{;\mu}\nu}\,\pi_{;\nu}^{\hphantom{;\nu}\rho}\,\pi_{;\rho}^{\hphantom{;\rho}\mu}\right)
+12 \left(\Box \pi\right) \left(\pi_{;\mu\nu}\pi^{;\mu\nu} \right)
+ 2 \left(\Box \pi\right) \left(\pi_{;\mu}\,\pi^{;\mu}\right) R\nonumber\\
&&+ 4 \left(\pi_{;\mu}\,\pi^{;\mu\nu}\,\pi_{;\nu}\right) R
+8 \left(\Box \pi\right) \left(\pi_{;\mu} \,R^{\mu \nu}\,\pi_{;\nu}\right)
- 4\left(\pi_{;\lambda}\,\pi^{;\lambda}\right)
\left(\pi_{;\mu\nu}\,R^{\mu \nu}\right)\nonumber \\
&&-16 \left(\pi_{;\mu}\,\pi^{;\mu\nu}\,R_{\nu\rho}\,\pi^{;\rho}\right)
- 8 \left(\pi_{;\mu}\,\pi_{;\nu}\,\pi_{;\rho\sigma}\,R^{\mu \rho \nu \sigma}\right).
\label{E4prime}
\end{eqnarray}
We see that this equation does not contain derivatives of order
higher than 2, and that it obviously reduces to the original form
of Ref.~\cite{Nicolis:2008in} in flat spacetime. On the other
hand, notice that it involves first-order derivatives of $\pi$ in
curved spacetime. This breaks the ``Galilean'' symmetry under
$\partial_{\mu} \pi \rightarrow \partial_{\mu} \pi + b_{\mu}$,
$\pi \rightarrow \pi + c$ which is a covariant generalization of
the transformation $\pi \rightarrow \pi + b_\mu x^\mu + c$ (where
$b_\mu$ and $c$ are constants) defined in
Ref.~\cite{Nicolis:2008in} in Minkowski spacetime. Note also the
complex mixing of the field degrees of freedom implied by the
presence of second derivatives of both $\pi$ and $g_{\mu\nu}$ in
this equation.

It is interesting to note that the full action $S_4 +
S^\text{nonmin}_4$, Eqs.~(\ref{S4}) and (\ref{Snm4}), can be
rewritten in a much more compact form thanks to integrations by
parts and commutations of derivatives:
\begin{equation} \label{S4tot}
S_4 + S^\text{nonmin}_4 =
\int d^4 x \sqrt{-g} \left(\pi_{;\lambda}\,\pi^{; \lambda}\right)
\left[2\left(\Box \pi\right)^2
- 2\left(\pi_{;\mu\nu}\,\pi^{;\mu\nu}\right)
- \frac{1}{2}\left(\pi_{;\mu}\,\pi^{;\mu}\right) R\right].
\end{equation}
The nonminimal coupling to the Ricci tensor (\ref{R2}) is indeed
automatically taken into account by the rewriting of
Eq.~(\ref{L4}) discussed below Eq.~(\ref{L5}).

Let us now investigate the variation of action $S_4$,
Eq.~(\ref{S4}), with respect to the metric\footnote{Note that
this variation was not computed in Ref.~\cite{Nicolis:2008in},
because it chose the $\pi$ energy-momentum tensor to be
negligible, as can be justified in an effective theory. Our aim is
here to exhibit the higher-derivatives it contains and to find a
cure, assuming it is a fundamental classical field theory.}.
It results in an
energy-momentum tensor\footnote{This definition of the
energy-momentum tensor corresponds to choosing units such that
$8\pi G/c^4 = 1$.} ${\cal T}^{\mu \nu}_4 \equiv (-g)^{-1/2}
\delta S_4/\delta g_{\mu\nu}$ of the form given by
\begin{eqnarray}
{\cal T}^{\mu \nu}_4 \label{TMN} &=&
\left(\pi^{;\mu}\,\pi^{;\nu}\right) \pi^{;\lambda}
\left(2\, \pi_{;\lambda\rho}^{\hphantom{;\lambda\rho}\rho}
- \pi^{;\rho}_{\hphantom{;\rho}\rho\lambda}\right)
\nonumber \\
&&+\left(\pi_{;\lambda}\,\pi^{;\lambda}\right)
\pi^{;\mu}
\left(\pi_{;\rho}^{\hphantom{;\rho}\rho\nu}
-\pi^{;\nu\rho}_{\hphantom{;\nu\rho}\rho}\right)
+\left(\pi_{;\lambda}\,\pi^{;\lambda}\right)
\pi^{;\nu}
\left(\pi_{;\rho}^{\hphantom{;\rho}\rho\mu}
-\pi^{;\mu\rho}_{\hphantom{;\mu\rho}\rho}\right)
\nonumber \\
&&-\pi^{;\lambda}\,\pi^{;\rho}
\left(\pi^{;\mu}\pi_{;\lambda\rho}^{\hphantom{;\lambda\rho}\nu}
+\pi^{;\nu}\pi_{;\lambda\rho}^{\hphantom{;\lambda\rho}\mu}\right)
+\left(\pi_{;\lambda}\,\pi^{;\lambda}\right)
\left(\pi_{;\rho}\,
\pi^{;\mu\nu\rho}\right) \nonumber \\
&&+\left(\pi_{;\lambda}\,\pi_{;\rho}\,\pi_{;\sigma}\,
\pi^{;\lambda\rho\sigma}\right)
g^{\mu \nu}
-\left(\pi_{;\lambda}\,\pi^{;\lambda}\right)
\left(\pi_{;\rho}\,\pi_{;\sigma}^{\hphantom{;\sigma}\sigma\rho}\right)
g^{\mu \nu} \nonumber \\
&&+\left(\pi^{;\mu}\,\pi^{;\nu}\right)
\left[3\left(\pi_{;\lambda\rho}\,\pi^{;\lambda\rho}\right)
-2\left(\Box \pi\right)^2\right]
+\left(\pi^{;\mu\nu}\right) \pi_{;\lambda}
\left(2\,\pi^{;\lambda\rho}\,\pi_{;\rho}
+ \pi^{;\lambda}\,\Box \pi\right)
\nonumber \\
&&+3\left(\Box \pi\right) \pi_{;\lambda}
\left(\pi^{;\lambda\mu}\,\pi^{;\nu}
+\pi^{;\lambda\nu}\,\pi^{;\mu}\right)
-4\,\pi_{;\lambda}\,\pi^{;\lambda\rho}
\left(\pi_{;\rho}^{\hphantom{;\rho}\mu}\,\pi^{;\nu}
+\pi_{;\rho}^{\hphantom{;\rho}\nu}\,\pi^{;\mu}\right)
\nonumber \\
&&-2 \left(\pi_{;\lambda}\,\pi^{;\lambda\mu}\right)
\left(\pi_{;\rho}\,\pi^{;\rho\nu}\right)
-\frac{1}{2} \left(\pi_{;\lambda}\,\pi^{;\lambda}\right)
\left[\left(\Box \pi\right)^2
+ \left(\pi_{;\rho\sigma}\,\pi^{;\rho\sigma}\right)\right]
g^{\mu \nu}
\nonumber \\
&&+ \pi_{;\lambda}\,\pi_{;\rho}
\left[3 \pi^{;\lambda\sigma}\,\pi_{;\sigma}^{\hphantom{;\sigma}\rho}
-2\left(\Box \pi\right) \pi^{;\lambda\rho}\right]
g^{\mu \nu}.
\end{eqnarray}
Notice that this energy-momentum tensor contains third-order
derivatives of $\pi$, and even if flat spacetime $g_{\mu\nu} =
\eta_{\mu\nu}$ were a solution of Einstein's equations. This
shows that once one lets the metric be dynamical, new degrees of
freedom will propagate even on a Minkowski background. One can
show that this energy-momentum tensor is conserved on shell.
Indeed, one finds (after many commutations of covariant
derivatives) that
\begin{eqnarray}
\nabla_\mu {\cal T}_{4}^{\mu \nu} =
\frac{1}{2}\,\pi^{;\nu}\,{\cal E}_4.
\label{conservTmunu}
\end{eqnarray}
This means in particular that the third derivatives present in
the expression of ${\cal T}_{4}^{\mu \nu}$ are killed by the
application of an extra covariant derivative.

Remarkably, it turns out that the addition to action (\ref{S4})
of the nonminimal coupling (\ref{Snm4}) is enough to eliminate
all third derivatives appearing in the $\pi$ energy-momentum
tensor. Indeed varying the sum of actions (\ref{S4}) and
(\ref{Snm4}), one finds now the energy-momentum tensor
\begin{eqnarray}
{\cal T}_{4}'^{\mu \nu} &=&
4
\left(\Box \pi\right)
\pi_{;\rho}
\bigl[\pi^{;\mu}\,\pi^{;\rho\nu}
+\pi^{;\nu}\,\pi^{;\rho\mu}\bigr]
- 2 \left(\Box \pi\right)^2 \left(\pi^{;\mu}\,\pi^{;\nu}\right)
+ 2 \left(\Box \pi\right)\left(\pi_{;\lambda}\,\pi^{;\lambda}\right)
\left(\pi^{;\mu\nu}\right)
\nonumber \\
&&+ 4
\left(\pi_{;\lambda}\,\pi^{;\lambda\rho}\,\pi_{;\rho}\right)
\left(\pi^{;\mu\nu}\right)
- 4 \left(\pi_{;\lambda}\,\pi^{;\lambda\mu}\right)
\left(\pi_{;\rho}\,\pi^{;\rho\nu}\right)
+ 2 \left(\pi_{;\lambda\rho}\,\pi^{;\lambda\rho}\right)
\left(\pi^{;\mu}\,\pi^{;\nu}\right)
\nonumber \\
&&- 2 \left(\pi_{;\lambda}\,\pi^{;\lambda}\right)
\left(\pi^{;\mu}_{\hphantom{;\mu}\rho}\,\pi^{;\rho\nu}\right)
- 4\,\pi^{;\lambda}\,\pi_{;\lambda\rho}
\bigl[\pi^{;\rho\mu}\,\pi^{;\nu}
+\pi^{;\rho\nu}\,\pi^{;\mu}\bigr]
-\left(\Box \pi\right)^2 \left(\pi_{;\lambda}\,\pi^{;\lambda}\right)
g^{\mu \nu}
\nonumber \\
&&- 4 \left(\Box \pi\right)
\left(\pi_{;\lambda}\,\pi^{;\lambda\rho}\,\pi_{;\rho}\right) g^{\mu \nu}
+ 4
\left(\pi_{;\lambda}\,\pi^{;\lambda\rho}\,\pi_{;\rho\sigma}\,\pi^{;\sigma}\right)
g^{\mu \nu}
\nonumber \\
&&+ \left(\pi_{;\lambda}\,\pi^{;\lambda}\right)
\left(\pi_{;\rho\sigma}\,\pi^{;\rho\sigma}\right) g^{\mu \nu}
+ \left(\pi_{;\lambda}\,\pi^{;\lambda}\right)
\left(\pi^{;\mu}\,\pi^{;\nu}\right) R
- \frac{1}{4}
\left(\pi_{;\lambda}\,\pi^{;\lambda}\right)
\left(\pi_{;\rho}\,\pi^{;\rho}\right)
g^{\mu \nu} R
\nonumber \\
&&- 2 \left(\pi_{;\lambda}\,\pi^{;\lambda}\right)
\pi_{;\rho}
\bigl[R^{\rho\mu}\,\pi^{;\nu}
+R^{\rho\nu}\,\pi^{;\mu}\bigr]
+ \frac{1}{2}
\left(\pi_{;\lambda}\,\pi^{;\lambda}\right)
\left(\pi_{;\rho}\,\pi^{;\rho}\right)
R^{\mu \nu}
\nonumber \\
&&+ 2 \left(\pi_{;\lambda}\,\pi^{;\lambda}\right)
\left(\pi_{;\rho}\,R^{\rho \sigma}\,\pi_{;\sigma}\right)
g^{\mu \nu}
-2 \left(\pi_{;\lambda}\,\pi^{;\lambda}\right)
\left(\pi_{;\rho}\,\pi_{;\sigma}\,R^{\mu\rho\nu\sigma}\right),
\end{eqnarray}
which now contains at most second derivatives. As expected, identity
(\ref{conservTmunu}) is also verified by ${\cal T}_{4}'^{\mu
\nu}$ and ${\cal E}'_4$, given in Eq.~(\ref{E4prime}) above.

As we will see now, things proceed along the same line for the
Lagrangian ${\cal L}_5$. Indeed, varying the action
\begin{eqnarray} \label{S5}
S_5=\int d^4 x \sqrt{-g}\, {\cal L}_5
\end{eqnarray}
with respect to $\pi$, we find, after commutation of covariant
derivatives, that the $\pi$ equations of motion do not contain
derivatives or order higher than 2 acting on $\pi$ but contain
third-order derivatives of the metric in the form of first
derivatives of the curvature. Those first derivatives are found
to be given by the combination
\begin{eqnarray}
&&-3 \left(\Box\pi\right) \left(\pi_{;\mu}\,\pi^{;\mu}\right)
\left(\pi_{;\nu}\,R^{;\nu}\right)
+ 3 \left(\pi_{;\mu}\,\pi^{;\mu\nu}\,\pi_{;\nu}\right)
\left(\pi_{;\rho}\,R^{;\rho}\right)
+ 6 \left(\Box\pi\right)
\left(\pi_{;\mu}\,\pi_{;\nu}\,\pi_{;\rho}\,R^{\mu\nu;\rho}\right)
\nonumber \\
&&+ 6 \left(\pi_{;\mu}\,\pi^{;\mu}\right)
\left(\pi_{;\nu}\,\pi_{;\rho\sigma}\,R^{\rho\sigma;\nu}\right)
- 12 \left(\pi_{;\mu}\,\pi_{;\nu}\,\pi^{;\rho}\,\pi_{;\rho\sigma}\,
R^{\mu\sigma;\nu}\right)
- 6 \left(\pi_{;\mu}\,\pi_{;\nu}\,\pi_{;\rho}\,\pi_{;\sigma\lambda}\,
R^{\mu\sigma\nu\lambda;\rho}\right).
\nonumber \\
\label{3ed}
\end{eqnarray}
One can think of eliminating those terms by adding to the action
a linear combination of the following seven non-trivial
contractions with the curvature tensors\footnote{These terms
are the only possible nonvanishing ones made of contractions
of five $\pi$'s (acted on by at least one derivative) and one
curvature tensor, involving a total of 8 derivatives.}:
\begin{subequations}
\begin{eqnarray}
{\cal L}_{5,1} &=& \left(\pi_{;\lambda}\,\pi^{;\lambda}\right)
\left(\pi_{;\mu}\, \pi_{;\nu}\, \pi_{;\rho\sigma}\, R^{\mu\rho\nu\sigma}\right),\\
{\cal L}_{5,2} &=& \left(\pi_{;\mu}\,\pi^{;\mu}\right)
\left(\pi_{;\nu}\,\pi^{;\nu}\right)
\left(\pi_{;\rho\sigma}\, R^{\rho\sigma}\right),\\
{\cal L}_{5,3} &=& \left(\pi_{;\mu}\,\pi^{;\mu}\right)
\left(\pi_{;\nu}\,\pi^{;\nu\rho}\,R_{\rho\sigma}\,\pi^{;\sigma}\right),\\
{\cal L}_{5,4} &=& \left(\pi_{;\mu}\,\pi^{;\mu}\right)
\left(\Box \pi\right)
\left(\pi_{;\nu} \,R^{\nu \rho}\,\pi_{;\rho}\right),\\
{\cal L}_{5,5} &=&
\left(\pi_{;\mu} \,\pi^{;\mu\nu}\,\pi_{;\nu}\right)
\left(\pi_{;\rho} \,R^{\rho\sigma}\,\pi_{;\sigma}\right),\\
{\cal L}_{5,6} &=& \left(\pi_{;\mu}\,\pi^{;\mu}\right)
\left(\pi_{;\nu}\,\pi^{;\nu}\right) \left(\Box \pi\right) R,\\
{\cal L}_{5,7} &=& \left(\pi_{;\mu}\,\pi^{;\mu}\right)
\left(\pi_{;\nu} \,\pi^{;\nu\rho}\,\pi_{;\rho}\right) R.
\end{eqnarray}
\end{subequations}
Not all those terms are independent, though. In fact using the
contracted Bianchi identity $R^{\mu\nu}_{\hphantom{\mu\nu};\nu} =
\frac{1}{2}\,R^{;\mu}$,
one can show that the combination ${\cal L}_{5,2}
+ 4{\cal L}_{5,3} - \frac{1}{2} {\cal L}_{5,6} - 2 {\cal L}_{5,7}$ is a
total derivative, and hence has an invariant action, i.e.,
\begin{eqnarray}
\delta\int d^4 x \sqrt{-g} \left({\cal L}_{5,2}
+ 4{\cal L}_{5,3} - \frac{1}{2} {\cal L}_{5,6}
- 2 {\cal L}_{5,7}\right) =0.
\end{eqnarray}
It turns out that there is a unique combination (up to the
addition of the above expression), which added to action $S_{5}$,
Eq.~(\ref{S5}), removes all higher derivatives (those of the
metric as well as those of $\pi$) in the $\pi$ equations of
motion, ${\cal E}'_5 = 0$. This combination is given by
\begin{eqnarray} \label{COMBI}
S^\text{nonmin}_5= \int d^4 x \sqrt{-g} \left(-3 {\cal L}_{5,1} - 18 {\cal L}_{5,3} + 3 {\cal L}_{5,4} + \frac{15}{2} \;{\cal L}_{5,7} \right).
\end{eqnarray}
The resulting field equation for $\pi$ is given in
Eq.~(\ref{Ep5}) of the appendix.

Similarly to Eq.~(\ref{S4tot}), the full action $S_5 +
S^\text{nonmin}_5$ may also be rewritten in various simpler forms
thanks to integrations by parts and commutations of derivatives.
A particularly elegant one is
\begin{eqnarray} \label{S5tot}
S_5 + S^\text{nonmin}_5 &=&
\frac{5}{2}\int d^4 x \sqrt{-g} \left(\pi_{;\lambda}\,\pi^{; \lambda}\right)
\bigl[\left(\Box \pi\right)^3
- 3 \left(\Box \pi\right) \left(\pi_{;\mu\nu}\,\pi^{;\mu\nu}\right)\nonumber \\
&&+2 \left(\pi_{;\mu}^{\hphantom{;\mu}\nu}\,
\pi_{;\nu}^{\hphantom{;\nu}\rho}\,\pi_{;\rho}^{\hphantom{;\rho}\mu}\right)
-6 \left(\pi_{;\mu}\,\pi^{;\mu\nu}\,G_{\nu\rho}\,\pi^{;\rho}\right) \bigr].
\end{eqnarray}

Varying now action (\ref{S5}) with respect to the metric, we find
as previously an energy-momentum tensor ${\cal T}_5^{\mu \nu}$
that contains third derivatives of $\pi$. However, it turns out
that the addition of (\ref{COMBI}) to (\ref{S5}) eliminates all
higher derivatives and generates a new energy-momentum tensor
${\cal T}_5'^{\mu \nu}$, whose expression is given in
Eq.~(\ref{Tp5}) of the appendix, and which contains at most
second derivatives (of $\pi$ and the metric). The analogue of
Eq.~(\ref{conservTmunu}) is consistently satisfied by
${\cal T}_5'^{\mu \nu}$ and ${\cal E}'_5$.

\section{Conclusions}
In this paper, we have shown that all higher-order derivatives
appearing in the field equations of the minimally-coupled
galileon to a dynamical metric, can be removed by a suitable
nonminimal coupling to curvature. This insures that no extra
degree of freedom is generated, and thereby defines a class
of purely scalar-tensor theories, involving a single scalar
degree of freedom, together with the standard graviton and
matter fields. However, note that the absence of higher
derivatives does not prove the stability of the theory (and
conversely, their presence may occur in some specific
stable models). This and other issues
deserve more investigation. For instance,
Ref.~\cite{Nicolis:2008in} considers Lagrangians involving
products of more that five $\pi$'s, which are total derivatives
in flat 4-dimensional spacetime, and it seems interesting to
study their behavior in curved and extradimensional manifolds.
Even without assuming the galileon symmetry of this reference, it
is worth studying the general form of scalar-field actions,
coupled to gravity, and yielding second-order equations. Sticking
with the original motivation of this kind of models, it remains
to study their precise phenomenological predictions and their
consistency in a cosmological context. We will tackle these
questions in a future study.

\begin{acknowledgments}
Our calculations have been cross-checked using several computer
programs, notably the \textit{xTensor} package~\cite{xTensor}
developed by J.-M.~Mart\'{\i}n-Garc\'{\i}a for
\textit{Mathematica}. It is a pleasure to thank G.~Dvali, J.~Mourad,
O.~Pujolas and I.~Sawicki for interesting discussions. The work
of A.V. was supported by the James Arthur Fellowship. A.V. would
like to thank the AstroParticule \& Cosmologie Laboratory for its
kind hospitality at the beginning of this project.
\end{acknowledgments}

\begin{appendix}
\section{Field equations deriving from the nonminimal extension
of ${\cal L}_5$}
We give below the field equation for $\pi$, ${\cal E}'_5 = 0$,
deriving from the action $S_5 + S^\text{nonmin}_5$, i.e., the sum
of Eqs.~(\ref{S5}) and (\ref{COMBI}):
\begin{eqnarray}
{\cal E}'_5&=&-5 \left(\Box\pi\right)^4
+30
\left(\Box\pi\right)^2 \left(\pi_{;\mu \nu}\,\pi^{;\mu \nu}\right)
+\frac{15}{2} \left(\Box\pi\right)^2 \left(\pi_{;\mu}\,\pi^{;\mu}\right) R
\nonumber \\
&&+15 \left(\Box\pi\right)^2 \left(\pi_{;\mu}\,R^{\mu \nu}\,\pi_{;\nu}\right)
-40\left(\Box\pi\right)
\left(\pi_{;\mu}^{\hphantom{;\mu}\nu}\,\pi_{;\nu}^{\hphantom{;\nu}\rho}\,\pi_{;\rho}^{\hphantom{;\rho}\mu}\right)
+15 \left(\Box\pi\right) \left(\pi_{;\mu}\,\pi^{;\mu \nu}\,\pi_{;\nu}\right) R
\nonumber \\
&&-30\left(\Box\pi\right) \left(\pi_{;\mu}\,\pi^{;\mu}\right)
\left(\pi_{;\nu \rho}\,R^{\nu \rho}\right)
-60 \left(\Box\pi\right)
\left(\pi_{;\mu}\,\pi^{;\mu\nu}\,R_{\nu \rho}\,\pi^{;\rho}\right)
\nonumber \\
&&-30 \left(\Box\pi\right) \left(\pi_{;\mu}\,\pi_{;\nu}\,\pi_{;\rho \sigma}\,R^{\mu \rho \nu \sigma}\right)
-15 \left(\pi_{;\mu \nu}\,\pi^{;\mu \nu}\right)
\left(\pi_{;\rho \sigma}\,\pi^{;\rho \sigma}\right)
\nonumber \\
&&+30 \left(\pi_{;\mu \nu}\,\pi^{;\nu \rho}\,\pi_{;\rho \sigma}\,\pi^{;\sigma\mu}\right)
-\frac{15}{2}
\left(\pi_{;\mu}\,\pi^{;\mu}\right) \left(\pi_{;\nu \rho}\,\pi^{;\nu \rho}\right) R
-15 \left(\pi_{;\mu}\,\pi^{;\mu\nu}\,\pi_{;\nu\rho}\,\pi^{;\rho}\right) R
\nonumber \\
&&-15 \left(\pi_{;\mu \nu}\,\pi^{;\mu \nu}\right)
\left(\pi_{;\rho}\,R^{\rho \sigma}\,\pi_{;\sigma}\right)
-30
\left(\pi_{;\mu}\,\pi^{;\mu \nu}\,\pi_{;\nu}\right)\left(\pi_{;\rho \sigma}\,R^{\rho \sigma}\right)
\nonumber \\
&&+30 \left(\pi_{;\mu}\,\pi^{;\mu}\right)
\left(\pi_{;\nu}^{\hphantom{;\nu}\rho}\,R_\rho^{\hphantom{\rho}\sigma}\,\pi_{;\sigma}^{\hphantom{;\sigma}\nu}\right)
+60 \left(\pi_{;\mu}\,\pi^{;\mu \nu}\,\pi_{;\nu\rho}\,R^{\rho \sigma}\,\pi_{;\sigma}\right)
\nonumber \\
&&+30 \left(\pi_{;\mu}\,\pi^{;\mu\nu}\,R_{\nu\rho}\,\pi^{;\rho\sigma}\,\pi_{;\sigma}\right)
+15 \left(\pi_{;\mu}\,\pi^{;\mu}\right)
\left(\pi_{;\nu \rho}\,\pi_{;\sigma \lambda}\,R^{\nu \sigma \rho \lambda}\right)
\nonumber \\
&&+30 \left(\pi_{;\mu}\,\pi_{;\nu}\,\pi_{;\rho \sigma}\,\pi^{;\sigma}_{\hphantom{;\sigma}\lambda}\,R^{\mu \rho \nu \lambda}\right)
-60 \left(\pi_{;\lambda}\,\pi^{;\lambda}_{\hphantom{;\lambda}\mu}\,\pi_{;\nu \rho}\,\pi_{;\sigma}\,R^{\mu\nu\rho\sigma}\right)
\nonumber \\
&&-\frac{15}{2}
\left(\pi_{;\mu}\,\pi^{;\mu}\right)
\left(\pi_{;\nu}\,R^{\nu \rho}\,\pi_{;\rho}\right) R
+15 \left(\pi_{;\mu}\,\pi^{;\mu}\right)\left(\pi_{;\nu}\,R^{\nu\rho}\,R_{\rho \sigma}\,\pi^{;\sigma}\right)
\nonumber \\
&&+15
\left(\pi_{;\mu}\,\pi^{;\mu}\right)
\left(\pi_{;\nu}\,\pi_{;\rho}\,R_{\sigma \lambda}\,R^{\nu \sigma \rho \lambda}\right)
-\frac{15}{2}
\left(\pi_{;\mu}\,\pi^{;\mu}\right)
\left(\pi_{;\nu}\,\pi_{;\rho}\,R^\nu_{\hphantom{\nu}\sigma\kappa\lambda}\,R^{\rho \sigma \kappa \lambda}\right).
\label{Ep5}
\end{eqnarray}
As mentioned in the bulk of the paper, both the galileon $\pi$
and the metric $g_{\mu\nu}$ are differentiated at most twice in
this field equation. This is also the case for the variation of
the same action $S_5 + S^\text{nonmin}_5$ with respect to
$g_{\mu\nu}$, i.e., the $\pi$ energy-momentum tensor, which takes
the form
\begin{eqnarray}
{\cal T}_{5}'^{\mu\nu}&=&-\frac{5}{2} \left(\Box\pi\right)^3
\left(\pi^{;\mu}\,\pi^{;\nu}\right)
-\frac{5}{2} \left(\Box\pi\right)^3
\left(\pi_{;\rho}\,\pi^{;\rho}\right)
g^{\mu \nu}
+\frac{15}{2} \left(\Box\pi\right)^2
\left(\pi_{;\rho}\,\pi^{;\rho}\right) \left(\pi^{;\mu \nu}\right)
\nonumber \\
&&+\frac{15}{2} \left(\Box\pi\right)^2
\pi_{;\rho}
\bigl[\pi^{;\rho\mu}\,\pi^{;\nu}
+\pi^{;\rho\nu}\,\pi^{;\mu}\bigr]
-\frac{15}{2} \left(\Box\pi\right)^2
\left(\pi_{;\rho}\,\pi^{;\rho \sigma}\,\pi_{;\sigma}\right)
g^{\mu \nu}
\nonumber \\
&&-15 \left(\Box\pi\right)
\left(\pi_{;\rho}\,\pi^{;\rho}\right)
\left(\pi^{;\mu\sigma}\,\pi_{;\sigma}^{\hphantom{;\sigma}\nu}\right)
+15 \left(\Box\pi\right)
\left(\pi_{;\rho}\,\pi^{;\rho \sigma}\,\pi_{;\sigma}\right)
\left(\pi^{;\mu \nu}\right)
\nonumber \\
&&+\frac{15}{2} \left(\Box\pi\right)
\left(\pi_{;\rho \sigma}\,\pi^{;\rho \sigma}\right)
\left(\pi^{;\mu}\,\pi^{;\nu}\right)
-15 \left(\Box\pi\right)
\left(\pi_{;\rho}\,\pi^{;\rho\mu}\right)
\left(\pi_{;\sigma}\,\pi^{;\sigma\nu}\right)
\nonumber \\
&&-15 \left(\Box\pi\right)
\pi^{;\rho}\,\pi_{;\rho\sigma}
\bigl[\pi^{;\sigma\mu}\,\pi^{;\nu}
+\pi^{;\sigma\nu}\,\pi^{;\mu}\bigr]
+\frac{15}{2} \left(\Box\pi\right)
\left(\pi_{;\rho}\,\pi^{;\rho}\right)
\left(\pi_{;\sigma \lambda}\,\pi^{;\sigma \lambda}\right)
g^{\mu \nu}
\nonumber \\
&&+15 \left(\Box\pi\right)
\left(\pi_{;\rho}\,\pi^{;\rho\sigma}\,
\pi_{;\sigma \lambda}\,\pi^{;\lambda}\right)
g^{\mu \nu}
+\frac{15}{4} \left(\Box\pi\right)
\left(\pi_{;\rho}\,\pi^{;\rho}\right)
\left(\pi^{;\mu}\,\pi^{;\nu}\right) R
\nonumber \\
&&-\frac{15}{2} \left(\Box\pi\right)
\left(\pi_{;\rho}\,\pi^{;\rho}\right)
\pi_{;\sigma}
\bigl[R^{\sigma\mu}\,\pi^{;\nu}
+R^{\sigma\nu}\,\pi^{;\mu}\bigr]
\nonumber \\
&&+\frac{15}{2} \left(\Box\pi\right)
\left(\pi_{;\rho}\,\pi^{;\rho}\right)
\left(\pi_{;\sigma}\,R^{\sigma \lambda}\,\pi_{;\lambda}\right)
g^{\mu \nu}
-\frac{15}{2} \left(\Box\pi\right)
\left(\pi_{;\rho}\,\pi^{;\rho}\right)
\left(\pi_{;\sigma}\,\pi_{;\lambda}\,
R^{\mu\sigma\nu\lambda}\right)
\nonumber \\
&&-\frac{15}{2}
\left(\pi_{;\rho}\,\pi^{;\rho}\right)
\left(\pi_{;\sigma \lambda}\,\pi^{;\sigma \lambda}\right)
\left(\pi^{;\mu \nu}\right)
+15 \left(\pi_{;\rho}\,\pi^{;\rho}\right)
\left(\pi^{;\mu\sigma}\,\pi_{;\sigma\lambda}\,\pi^{;\lambda\nu}\right)
\nonumber \\
&&-15 \left(\pi_{;\rho}\,\pi^{;\rho \sigma}\,\pi_{;\sigma}\right)
\left(\pi^{;\mu\lambda}\,\pi_{;\lambda}^{\hphantom{;\lambda}\nu}\right)
-15 \left(\pi_{;\rho}\,
\pi^{;\rho\sigma}\,
\pi_{;\sigma \lambda}\,
\pi^{;\lambda}\right)
\left(\pi^{;\mu \nu}\right)
\nonumber \\
&&-5
\left(\pi_{;\rho}^{\hphantom{;\rho}\sigma}\,
\pi_{;\sigma}^{\hphantom{;\sigma}\lambda}\,
\pi_{;\lambda}^{\hphantom{;\lambda}\rho}\right)
\left(\pi^{;\mu}\,\pi^{;\nu}\right)
-\frac{15}{2}
\left(\pi_{;\sigma \lambda}\,\pi^{;\sigma \lambda}\right)
\pi_{;\rho}
\bigl[\pi^{;\rho\mu}\,\pi^{;\nu}
+\pi^{;\rho\nu}\,\pi^{;\mu}\bigr]
\nonumber \\
&&+15\,\pi_{;\rho}\,
\pi^{;\rho \sigma}\,
\pi_{;\sigma \lambda}
\bigl[\pi^{;\lambda\mu}\,\pi^{;\nu}
+\pi^{;\lambda\nu}\,\pi^{;\mu}\bigr]
+15\,\pi^{;\rho}\,\pi_{;\rho\lambda}\,\pi_{;\sigma}
\bigl[\pi^{;\lambda\mu}\,\pi^{;\sigma\nu}
+\pi^{;\lambda\nu}\,\pi^{;\sigma\mu}\bigr]
\nonumber \\
&&-5 \left(\pi_{;\rho}\,\pi^{;\rho}\right)
\left(\pi_{;\sigma}^{\hphantom{;\sigma}\lambda}\,
\pi_{;\lambda}^{\hphantom{;\lambda}\kappa}\,
\pi_{;\kappa}^{\hphantom{;\kappa}\sigma}\right)
g^{\mu \nu}
+\frac{15}{2}
\left(\pi_{;\rho}\,\pi^{;\rho \sigma}\,\pi_{;\sigma}\right)
\left(\pi_{;\lambda \kappa}\,\pi^{;\lambda \kappa}\right)
g^{\mu \nu}
\nonumber \\
&&-15 \left(\pi_{;\rho}\,\pi^{;\rho\sigma}\,\pi_{;\sigma\lambda}\,
\pi^{;\lambda\kappa}\,\pi_{;\kappa}\right)
g^{\mu \nu}
-\frac{15}{4}
\left(\pi_{;\rho}\,\pi^{;\rho}\right)
\pi_{;\sigma}
\bigl[\pi^{;\sigma\mu}\,\pi^{;\nu}
+\pi^{;\sigma\nu}\,\pi^{;\mu}\bigr] R
\nonumber \\
&&+\frac{15}{4} \left(\pi_{;\rho}\,\pi^{;\rho}\right)
\left(\pi_{;\sigma}\,\pi^{;\sigma \lambda}\,\pi_{;\lambda}\right)
R\,g^{\mu \nu}
-\frac{15}{2} \left(\pi_{;\rho}\,\pi^{;\rho}\right)
\left(\pi_{;\sigma}\,\pi^{;\sigma \lambda}\,\pi_{;\lambda}\right)
R^{\mu \nu}
\nonumber \\
&&-\frac{15}{2} \left(\pi_{;\rho}\,\pi^{;\rho}\right)
\left(\pi_{;\sigma}\,R^{\sigma \lambda}\,\pi_{;\lambda}\right)
\left(\pi^{;\mu \nu}\right)
-\frac{15}{2} \left(\pi_{;\rho}\,\pi^{;\rho}\right)
\left(\pi_{;\sigma\lambda}\,R^{\sigma\lambda}\right)
\left(\pi^{;\mu}\,\pi^{;\nu}\right)
\nonumber \\
&&+\frac{15}{2} \left(\pi_{;\rho}\,\pi^{;\rho}\right)
\pi^{;\sigma}\,\pi_{;\sigma\lambda}
\bigl[R^{\lambda\mu}\,\pi^{;\nu}
+R^{\lambda\nu}\,\pi^{;\mu}\bigr]
\nonumber \\
&&+\frac{15}{2} \left(\pi_{;\rho}\,\pi^{;\rho}\right)
\pi_{;\lambda}\,\pi_{;\sigma}
\bigl[R^{\lambda\mu}\,\pi^{;\sigma\nu}
+R^{\lambda\nu}\,\pi^{;\sigma\mu}\bigr]
\nonumber \\
&&+\frac{15}{2} \left(\pi_{;\rho}\,\pi^{;\rho}\right)
\pi^{;\sigma}\,R_{\sigma \lambda}
\bigl[\pi^{;\lambda\mu}\,\pi^{;\nu}
+\pi^{;\lambda\nu}\,\pi^{;\mu}\bigr]
-15
\left(\pi_{;\rho}\,\pi^{;\rho}\right)
\left(\pi_{;\sigma}\,\pi^{;\sigma\lambda}\,
R_{\lambda\kappa}\,\pi^{;\kappa}\right)
g^{\mu \nu}
\nonumber \\
&&+\frac{15}{2} \left(\pi_{;\rho}\,\pi^{;\rho}\right)
\pi_{;\sigma}\,\pi_{;\lambda \kappa}
\bigl[R^{\mu\lambda\sigma\kappa}\,\pi^{;\nu}
+R^{\nu\lambda\sigma\kappa}\,\pi^{;\mu}\bigr]
\nonumber \\
&&-\frac{15}{2} \left(\pi_{;\rho}\,\pi^{;\rho}\right)
\pi_{;\sigma}\,\pi_{;\lambda}
\bigl[R^{\mu\sigma \lambda \kappa}\,
\pi_{;\kappa}^{\hphantom{;\kappa}\nu}
+ R^{\nu\sigma \lambda \kappa}\,
\pi_{;\kappa}^{\hphantom{;\kappa}\mu}\bigr]
\nonumber \\
&&+\frac{15}{2} \left(\pi_{;\rho}\,\pi^{;\rho}\right)
\pi^{;\sigma}\,\pi_{;\sigma\lambda}\,\pi_{;\kappa}
\bigl[R^{\mu\lambda\nu\kappa}
+R^{\nu\lambda\mu\kappa}\bigr]
\nonumber \\
&&-\frac{15}{2} \left(\pi_{;\rho}\,\pi^{;\rho}\right)
\left(\pi_{;\sigma}\,\pi_{;\lambda}\,\pi_{;\kappa\tau}\,
R^{\sigma\kappa\lambda\tau}\right)
g^{\mu \nu}.
\label{Tp5}
\end{eqnarray}

\end{appendix}
\end{document}